\documentclass[12pt]{article}
\usepackage{amssymb,amsmath,esint,titlesec}
\titleformat*{\section}{\large\bf}
\titleformat*{\subsection}{\normalsize\bf}

\begin{document}


\begin{titlepage}

                               \vspace*{0mm}

                             \begin{center}

\Large\bf  Convexity and the Euclidean metric of space-time

                            \vspace*{3.5cm}

              \normalsize\sf    NIKOLAOS  \  KALOGEROPOULOS$^\dagger$\\

                            \vspace{0.2cm}
                            
 \normalsize\sf Carnegie Mellon University in Qatar\\
 Education City,   P.O.  Box 24866\\
 Doha,  Qatar\\

                            \end{center}

                            \vspace{3.5cm}

                     \centerline{\normalsize\bf Abstract}
                     
                           \vspace{3mm}
                     
\normalsize\rm\setlength{\baselineskip}{18pt} 

We address the question about the reasons why the ``Wick-rotated", positive-definite, space-time metric obeys the 
Pythagorean theorem.  An answer is proposed  based  on the convexity and smoothness properties of the functional 
spaces purporting to provide the kinematic framework of approaches to quantum gravity. We employ moduli of convexity 
and smoothness which are eventually extremized by Hilbert spaces. We point out the potential physical significance that 
functional analytical dualities play in this framework. Following the spirit of the variational principles employed in classical 
and quantum Physics, such Hilbert spaces dominate  in a generalized functional integral approach.  
The metric of space-time is induced by the inner product of such Hilbert spaces.\\

                           \vfill

                                                                         
                             \vfill

\noindent\rule{13cm}{0.2mm}\\  
   \noindent $^\dagger$ \small\rm Electronic addresses: \ \  {\normalsize\sf nkaloger@cmu.edu, \ \  nikos.physikos@gmail.com}\\

\end{titlepage}
 

                                                                                \newpage                 

\rm\normalsize
\setlength{\baselineskip}{18pt}

                                              \section{Introduction.}

When one looks at the equations describing the four fundamental interactions of nature, then s/he immediately notices 
that the kinematic equations in the Lagrangian formalism involve second order derivatives with respect to space-time variables. 
It may be worthwhile to try to understand the reasons why, something which is usually taken for granted as an empirical fact
since the earliest days of Newtonian mechanics.   
 Modelling of the fundamental interactions except gravity relies, at the classical level, on Classical Mechanics, on  Electromagnetic Theory and its Yang-Mills/non-abelian gauge 
``generalizations". General Relativity  can also be seen as a gauge theory whose gauge group is the diffeomorphism group (re-parametrization invariance) 
of the underlying topological space endowed with its metric structure.\\ 

In all of the above, and in the Lagrangian approach which we employ throughout this work, the Euler-Lagrange equations that describe the underlying dynamics 
can be seen to emerge from variational principles; such equations could use derivatives of arbitrarily high order and a formalism for accommodating this fact 
has already been developed.  However, in practice, when dealing with fundamental interactions and not performing perturbative or approximate calculations that 
rely on series expansions,  one rarely needs derivatives that are of higher than the second-order with respect to space-time variables.\\

The statements on the number of derivatives in the equations of dynamics can be seen to be essentially equivalent, 
upon partial integration, to the fact that the kinetic terms as well as relevant 
potential energy terms are  at most quadratic with respect to first order derivatives of the fundamental variables/fields/order parameters. 
This in turn, allows one to use Euclidean/Riemannian concepts to model the evolution of such systems; for particle systems one uses the more familiar aspects of 
finite-dimensional  Riemannian spaces \cite{Pettini}, and for  field theories one may have to resort to using aspects of infinite dimensional  
manifolds \cite{KM} which involve further subtleties.\\   
 
One could use, equivalently in the simplest context, a Hamiltonian approach where the equations involve of first order derivatives. 
Each of these two lines of approach Lagrangian versus Hamiltonian has its own advantages and drawbacks, as is well-known. 
See, for instance, the very recent review \cite{Cor} for an approach to gravity based on first order actions including boundary terms.   
Without denying the advantages of the Hamiltonian or first order formalisms,  we will adopt in this work the Lagrangian/second-order formalism, as already mentioned above. 
Our viewpoint is somewhat influenced by and may have common points with that of \cite{KarRaj}, even though the methods which we use 
and the results we reach are substantially different from those of that work.\\   

We will  adopt the convenient technical device of Wick-rotating to a ``Euclidean"  (i.e. positive-definite) signature metric.
We would like to state at this point, that we do not consider positive-definite metrics ``superior" in any sense, to these of 
Lorentzian (indefinite) signature. On the contrary, all experiments, so far,  point to the latter as being the physical ones, so even if unstated, our 
underlying view is closer to  \cite{Sorkin1} rather than \cite{Bojo1}, for instance.  On the other hand, on purely formalistic grounds,
 the Euclidean (positive-definite) metrics  are easier to work with, as they obey the positivity 
property and the triangle inequality that their indefinite metric (Lorentzian) counterparts are lacking. \\
 
A substantial amount of effort has been spent, in recent decades, into understanding the non-trivial features of space-time, as described by
the General Theory of Relativity or other theories incorporating space-time diffeomorphism invariance.  Even though considerable progress 
has been made toward such an understanding, it is probably fair to state that many important issues still remain  unresolved \cite{NVis}. 
It is not clear, for instance, why space-time is 4-dimensional, to what extent it is smooth and how such a smoothness arises,  why its Wick-rotated 
metric obeys the Pythagorean theorem etc. Henceforth we will work only with positive definite (Wick-rotated) metrics of space-time 
and by ``Euclidean" we will mean only the ones obeying the Pythagorean theorem.
In linear algebraic and functional analytic language the metrics we call Euclidean would be called the $l^2$ metrics. Moreover, we will use the term 
space-time in order to keep in mind that our arguments actually purport to describe (indefinite/Lorentzian signature) space-time 
even if we use positive-definite (Wick-rotated) metrics throughout this work.  \\

A potentially fruitful way toward answering why the space-time metric is Euclidean, which is the subject of this work, 
is to look at it through the eyes of convexity. Convexity plays a central role in many branches of Mathematics, but  seems to be under-appreciated 
and under-utilized in gravitational Physics \cite{GibIsh}, and not only. This can also be considered as a partial motivation for looking for answers
to the questions of our interest through convexity. In the context of linear spaces, convexity turns out to be dual to smoothness, a fact that we also 
use to support the case for the Euclidean form of the space-time metric.  \\

In Section 2, we provide a general background and a physical interpretation, wherever feasible, from the theory of normed spaces with emphasis 
on properties pertinent to our arguments.  \ In Section 3, we discuss the aspects of convexity, smoothness and their duality via a Legendre-Fenchel 
transform, employing in particular Clarkson's modulus of convexity and to the Day-Lindenstrauss modulus of smoothness.  
In Section 4, we put our, less than rigorous, argument together on how the previous results result in space-time metric that has the Euclidean form.
Section 5, presents some conclusions and caveats.\\
   

                                                        \section{Background and  physical  interpretation of some concepts.}
                                               
Space-times are assumed to be locally flat, to first order approximation of their metric. 
This is an outcome of the application of the Equivalence Principle, 
and a well-known fact in Riemannian/Lorentzian geometry.   
Therefore, we can analyze the ultralocal aspects of the features of a space-time  by confining our attention to vector spaces. 
Most features of such tangent spaces can be captured by the  various closely related functional spaces that can be defined on them.
There are numerous classes of functional spaces that have been investigated during the last century, in the context of Functional Analysis,
such as Lebesgue, Hardy, Bergman, Sobolev, Orlicz, Besov, Triebel-Lizorkin, etc spaces. Most of the previously named spaces are or rely on, in one form 
or another,  constructions and results of (usually infinite dimensional) Banach spaces. Such infinite dimensional Banach spaces will be the main vector 
spaces of interest in this work. \\

 In this Section we provide some preliminary information on these and related mathematical constructions which are 
pertinent to the subsequent sections, where most of our arguments are developed. 
We attempt to stress their physical motivation and interpretation, wherever feasible, from the viewpoint we adopt in this work.  \\


\subsection{Norms on linear spaces.}

To make the exposition more readable and self-contained,  we recall \cite{Rudin, LiebLoss} that a norm on a vector space $\mathcal{V}$ 
defined over a field $\mathbb{F}$, usually \  $\mathbb{F} = \mathbb{R}$ \ or \ $\mathbb{F} = \mathbb{C}$ \ in most applications in Physics, 
is a function \ $\| \cdot \| : \mathcal{V} \rightarrow \mathbb{R}_+$ \  where \ $\mathbb{R}_+ = \{ \lambda \in\mathbb{R} : \lambda \geq 0 \}$, \ 
such that for all \  $x, y \in \mathcal{V}$
\begin{itemize}
   \item Positive definiteness: \  $\| x \| \ = \ 0$,  \ \ if and only if \ \  $x \ = \ 0$.
   \item Homogeneity: \hspace{12mm} $ \| k x\| \ = \ |k| \| x \| $, \ for all \ $x\in \mathcal{V}$, \  and all \  $k\in \mathbb{F}.$  
   \item Triangle inequality: \ \   $\| x+ y \| \ \leq \ \| x \| + \| y \|$
\end{itemize}
A vector space $\mathcal{V}$ endowed with a norm, which moreover is  complete in this norm, namely such that all Cauchy sequences have limits belonging to 
$\mathcal{V}$, is called a Banach space. Examples of Banach spaces that are explicitly used in this work are:
\begin{itemize}
       \item $c_0$ \ is the space of all sequences $a = (a_n), \ n\in\mathbb{N}$ converging to zero,  with \ $a_n \in \mathbb{F}$, \ and the sup-norm    
                   \begin{equation}
                          \| a \| \ = \ \sup_n |a_n |
                   \end{equation}
       \item   $l^p, \ 1\leq p < \infty$, \ which is the space of all sequences $a = (a_n), \ n\in \mathbb{N}$ with $a_n \in \mathbb{F}$ endowed with the norm 
                 \begin{equation}  
                       \| a \|_p = \left( \sum_{i=1} ^n |a_n|^p \right)^\frac{1}{p} < \infty
                 \end{equation}
      \item   $l^\infty$, the space of all bounded sequences \ $a = (a_n), \ n\in\mathbb{N}$ \ endowed with the supremum norm 
                 \begin{equation}
                           \| a \| \ = \ \sup_n |a_n |
                 \end{equation}     
     \item $L^p (\mathbb{R}^n), \ 1 \leq p < \infty$, \  the space of Lebesgue integrable functions \ $f: \mathbb{R}^n \rightarrow \mathbb{F}$ \ endowed with the norm
                 \begin{equation}
                          \| f \|_p \ = \ \left( \int_{\mathbb{R}^n} |f|^p \right)^\frac{1}{p} 
                 \end{equation}  
    \item $L^\infty (\mathbb{R}^n)$, the space of all essentially bounded $f: \mathbb{R}^n \rightarrow \mathbb{F}$, endowed with the norm
                \begin{equation}
                      \| f \|_\infty \ = \ \inf \{ C: \  |f| < C \ \ \mathrm{ almost \ everywhere} \}
                \end{equation}
\end{itemize}
The above functional spaces \ $L^p (\mathbb{R}^n), \ 1\leq p \leq \infty$ \ are, strictly speaking, spaces over equivalence classes of functions, where  two functions 
are considered equivalent (``equal") if they differ from each other, at most, in a set of measure zero. One uses the Lebesgue measure 
(``volume") of $\mathbb{R}^n$ in the definition of such $L^p(\mathbb{R}^n)$. In most applications in Physics, the distinction between functions and classes of 
equivalent functions is tacitly assumed and not explicitly stated. For completeness, we mention that a Hilbert space is a linear space endowed with an inner product 
$(\cdot, \cdot)$ which is, moreover, complete \cite{Rudin, LiebLoss}. The norm which we will assume that Hilbert spaces are endowed with,
is induced by their inner product by \ $\| x\|^2 \ = \ (x, x)$. \ Among the above examples of Banach spaces, \ $l^2$ \ and \ $L^2(\mathbb{R}^n)$ \ are Hilbert spaces.  
Since the spaces that we will be referring to are normed, hence  metrizable, they are endowed with the topology induced by the metric. An important topological 
property of such topological (vector) spaces  is  separability: this means that such spaces contain a countable dense subset. For such metrizable spaces, 
being separable is equivalent to being second countable. Topological spaces that are finite or countably  infinite are, obviously, separable. For Hilbert spaces 
separability implies the existence of a countable orthonormal basis, hence any separable infinite dimensional Hilbert space is isometric to $l^2$. In most cases in
quantum theory, separability of the space of the Hllbert space of wave-functions is tacitly assumed. Banach spaces can also be either separable or not separable \cite{HSVZ}. 
The issue of  separabilty of the spaces we use is not pertinent to the arguments of this work, so it will not be encountered anywhere in the sequel.\\


\subsection{Norm equivalence.} 

Naturally, one can endow a vector space $\mathcal{V}$ with many different norms. Usually the ``appropriate" choice of such a norm has substantial implications 
for the specific predictions of the physical model built on $(\mathcal{V}, \| \cdot \|)$. In other words, the choice of two ``different", in the naive sense of the word, 
norms will usually result in substantially different predictions of the physical quantities resulting form such calculations. Hence the choice of a norm is usually
considered to be a piece of data which is initially provided by hand, in any model. In most cases such a choice of norm is not explicitly discussed, because it is 
assumed that a norm arising from an inner product is the physically relevant one. In this work, we would like to know why this may be the case.\\

Given that a vector space may be endowed with different norms, one can ask what the word ``different" may actually mean. If two norms are point-wise different but still 
reasonably close to each other, in some particular sense, let's say with respect to a particular metric, can they still be declared as ``equivalent"? Consider, for instance,
the case of Hamiltonian systems of many degrees of freedom. This paragraph operates in the context of Euclidean metrics but can still be used to highlight our 
point of view. Suppose that one changes the metric of the phase space. If all thermodynamic quantities 
remain invariant under such a change of norm, is it reasonable to consider the two metrics as ``equivalent" or should someone insist as treating them as ``different"?
Under some additional conditions, two such metrics may turn out to be equal; this is reminiscent of the (Hamburger, Stieltjes etc) moment problem pertaining to the 
equality of underlying probability distributions, if the sets of moments of these distributions are equal \cite{Simon}. Something similar occurs in 
Geometry  or Analysis: quite often two metrics or norms are declared as ``equivalent" even though they may be pointwise distinct. \\

This re-interpretation of the concept of ``equivalence" can have substantial consequences for Statistical or Quantum Physics. The issue at hand can be seen as a form 
of  ``stability". Consider, for instance, a Hamiltonian system. This determines a symplectic structure on the phase space $\mathcal{M}$ of the system or, alternatively, 
a Poisson structure on the  space of smooth functions $C^\infty (\mathcal{M})$ \cite{Weinstein}. However, it does not generically determine a metric structure on 
$\mathcal{M}$. Such a metric is usually assumed to stem from the quadratic ``kinetic" term of the Hamiltonian, at least for  systems having a kinetic term of such 
form. Then we can  proceed to perform an analysis of the evolution of this Hamiltonian system, quantize the system etc. based on this symplectic and metric structures.  
However, one should not forget that the metric structure was not ``natural", or may not be unique. For this reason if it changes, especially a little bit, one would expect the 
physically relevant results to remain practically invariant. This can be interpreted as a form of structural stability of the underlying Hamiltonian dynamical system, even though 
it applies to an auxiliary piece of data, such as the metric. It is a subject of much discussion on whether such a metric structure 
should be assumed, and if so, to what extent it determines the statistically significant features of the system in an appropriate many-body ``thermodynamic" limit 
\cite{Klauder}. \\

Given the above considerations, one may be willing to allow variations of the assumed metric of the underlying Hamiltonian system. Then two metrics can be declared as 
equivalent, as also previously mentioned, if they give the same predictions of physically relevant quantities. The question is then how to find such variations of metrics, or 
norms. From an analytical viewpoint, one can use a central concept of Analysis, that of the limit, to determine how to answer such questions. The most crude/rough approach 
is to demand that two metrics/norms \ $\| \cdot \|_1, \ \| \cdot \|_2$ \ to be considered as equivalent if all sequences converging with respect to one of them also converge with 
respect to the other. This is realized when there are constants \ $0 < C_1 \leq C_2 $ \ such that                 
\begin{equation}
          C_1 \| \cdot \|_1 \ \leq \ \|  \cdot \|_2 \ \leq \  C_2 \| \cdot \|_1      
\end{equation}
This always holds for any two norms on finite dimensional vector spaces. However, it is not true in general for infinite dimensional functional spaces, which are at the center 
of our attention.  \\

A way to interpret (6) from the viewpoint of statistical theories is as follows: physically relevant results of the microscopic or quantum dynamics should be somehow reflected  
or emerge in the large scale/multi-particle or thermodynamic limit. Such features should remain largely unaffected by most ``small-scale" details of the system or their 
perturbations. 
This tacitly assumes that the underlying quantities characterizing geometric characteristics of the system are proportional to the effective measure(s) used in the  
calculations of the pertinent statistical quantities. This is clearly true for ergodic systems, but it can also be true for non-ergodic systems such as the ones 
whose thermodynamic behavior is conjecturally described by any of the many recently proposed entropic functionals, such as  the ``Tsallis entropy" \cite{T-book}, or 
the ``$\kappa$-entropy" \cite{Kan},  for instance.  
So, from a geometric viewpoint, equations like (6) express this insensitivity to small-scale details. 
This viewpoint motivates and pervades ``coarse geometry" \cite{Roe} and is frequently encountered in constructions 
related to hyperbolic spaces \cite{BridHaef} or groups \cite{Gromov1}.  Metrically, it is expressed by demanding invariance under quasi-isometries. 
In dynamical systems one can see a similar viewpoint in several occasions, an example of which is that the topological entropy of a map or flow on a metric 
space $(\mathfrak{X}, d)$ does not actually depend on the specific metric/distance  function  $d$, but only on the class on metrics on $\mathfrak{X}$ that induce 
the same topology on $\mathfrak{X}$. Then the key/desired  invariance akin to (6), is the topological conjugacy \cite{KatokHas}. \\       


\subsection{The operator norm and the Banach-Mazur distance.}

 Before continuing, for completeness of the exposition,  we state two definitions that will be extensively used in the sequel \cite{Rudin, LiebLoss}. 
Let $(\mathcal{X}, \| \cdot \|_\mathcal{X} )$ and $(\mathcal{Y}, \| \cdot \|_\mathcal{Y} )$ be two normed spaces, over $\mathbb{R}$, $\mathbb{C}$ (or any other field, 
although the general case does not appear to be of any particular interest in Physics, so far)
and let $T$ be a continuous linear map $T:\mathcal{X} \rightarrow \mathcal{Y}$.
If such a map exists between $\mathcal{X, Y}$ which is bijective, and its inverse $T^{-1}$ is  also bijective, then $\mathcal{X, Y}$ are called isomorphic. 
Actually less is needed: the Open Mapping Theorem guarantees that if $T$ is bijective and bounded, so is $T^{-1}$.
If a mapping $T$ is an isometry, namely if 
\begin{equation}
   \| Tx \|_\mathcal{Y} \ = \ \| x \|_\mathcal{X}
\end{equation} 
then $\mathcal{X, Y}$ are called isometric. Since boundedness of $T$ is equivalent to its continuity, one can see that isometric spaces are isomorphic.  
Let the space of bounded linear maps from $\mathcal{X}$ to $\mathcal{Y}$ be denoted by $\mathcal{B} (\mathcal{X}, \mathcal{Y})$. This space
$\mathcal{B}(\mathcal{X}, \mathcal{Y})$ can be endowed with the operator (sup-) norm which for $T\in \mathcal{B}(\mathcal{X}, \mathcal{Y})$ 
is given by the following equivalent definitions 
\begin{equation}  
    \| T \| \ = \  \sup_{x\in\mathcal{X}} \ \{\| Tx \| : \| x \| \leq 1 \} \ = \ \sup_{x\in\mathcal{X}} \  \{ \| Tx \| : \| x \| = 1 \} \ =
                                                  \ \sup_{x\in\mathcal{X}} \ \left\{ \frac{\| Tx \|}{\| x \|} : \| x \| \neq 0 \right\}
\end{equation} 
It turns out that if $\mathcal{X}$ is a normed space and if $\mathcal{Y}$ is a Banach space, then $\mathcal{B} (\mathcal{X}, \mathcal{Y})$ endowed 
with the operator norm (8) is a Banach space too. 
We know that every normed space can be isometrically embedded in a Banach space. So, most of the pertinent features of general normed 
spaces are contained in Banach spaces, therefore we can use the latter believing that we are not losing important aspects  of the flexibility or generality of 
the former, for applications in Physics. Given this, our question is then reduced to  asking,  why among all Banach spaces, the inner product (Hilbert) spaces 
are the ones describing most fundamental aspects of nature most accurately, so far as we know today. \\

In the spirit of norm-equivalence, discussed in Subsection 2.2, one can ask how close, or how far, from each other are two linear spaces. 
From a metric viewpoint they are identical, if they are isometric. We want to have a ``reasonable" distance function that measures how far from each other 
they may be, if they are not isometric.
Defining such a distance function is clearly a matter of choice which ideally, for our purposes, should also reflect some desirable physical properties. 
It appears that the classical Banach-Mazur distance has properties fitting such requirements. It is defined as follows. Let $\mathcal{X}, \mathcal{Y}$ be two 
isomorphic Banach spaces. The Banach-Mazur distance between them is defined as 
\begin{equation}    
   d_{BM} (\mathcal{X}, \mathcal{Y}) \ = \ \inf_T  \  \{ \| T \| \cdot \| T^{-1}\| \}
\end{equation}
where \ $T: \mathcal{X} \rightarrow \mathcal{Y}$ \  is an isomorphism. If $\mathcal{X}, \mathcal{Y}$ are not isomorphic, then their Banach-Mazur distance is infinite, 
by definition.  We can immediately see that 
\begin{equation}
     d_{BM} (\mathcal{X}, \mathcal{Y}) \geq 1
\end{equation}
and that for isometric spaces such a distance is exactly equal to one. The converse is also true, but only for finite-dimensional Banach spaces. 
Therefore, the Banach-Mazur distance is actually a distance function on the set of equivalence classes of normed spaces (where ``equivalence" is defined as ``isometry"), 
but in a multiplicative sense, namely for three isomorphic linear spaces \ $\mathcal{X}, \ \mathcal{Y}, \ 
\mathcal{Z}$, \  it satisfies 
\begin{equation} 
    d_{BM} (\mathcal{X}, \mathcal{Z}) \ \leq \  d_{BM} (\mathcal{X} , \mathcal{Y}) \cdot d_{BM} (\mathcal{Y}, \mathcal{Z}) 
\end{equation}
To get to the usual triangle inequality instead of (11), we have to consider the logarithm of $d_{BM}$. \ The Banach-Mazur distance is 
invariant under invertible linear maps $T$, namely 
\begin{equation}
    d_{BM} (T\mathcal{X}, T\mathcal{Y}) \ = \ d_{BM} (\mathcal{X}, \mathcal{Y})
\end{equation}
In some sense the Banach-Mazur distance expresses the minimum distortion that any isomorphism between two linear spaces can possibly entail. \\ 
 
Before closing this Subsection, one cannot help but distinguish between the functional spaces arising in a theory developed on space-time and the underlying form of the 
metric of space-time itself. These are clearly two quite distinct classes of spaces that have to be treated independently. However, one can hope that 
if and when a reasonably testable model  of quantum gravity is found, then its Hilbert (or more generally, functional) space of ``wave-functions" will induce 
the observable  metric of space-time. So from now on, the working assumption will be that such quantum mechanical Hilbert spaces induce 
the space-time metric. Therefore we should  address the question about what is so special about inner product (Hilbert) spaces among the class of all Banach spaces.
Since we will be working with spaces of functions, we will focus on infinite dimensional Banach spaces in the sequel.
To be more concrete, we will have in mind spaces seen frequently in applications such as the spaces of $p$-summable sequences ($l^p$), or of Lebesgue $p$-integrable functions 
on $\mathbb{R}^n$ ($L^p$) endowed with their cardinality or their  induced Euclidean measure ``volume" respectively, appearing in (1)-(5). \\  


\subsection{Reflexive and super-reflexive spaces.} 

To proceed in determining desirable properties of the Banach spaces of functions on $\mathbb{R}^n$, we consider the following. It is widely believed among many,
or even most, quantum gravity practitioners that spacetime properties such as its topology, smoothness, metric etc  should be ``derived" from a quantum theory of 
gravity rather than be put in the models by hand. 
This seems to be a widespread belief, regardless of the exact approach to quantum gravity that someone follows. It is based in the fact that        
macroscopic properties of systems can be derived from their quantum counterparts and rely, to some extent, in the great separation of scales and numbers of 
degrees of freedom between the microscopic and the macroscopic scales. This is the main reason, as accepted today, of why Statistical Mechanics works so 
well in providing accurate predictions for systems with many degrees of freedom. \\

Independently of such physical considerations, there has been a recent surge of activity in
Geometry and Analysis purporting to better understand first order calculus properties of non-smooth spaces \cite{Hein}. An  influential work in this direction has 
been \cite{Cheeger}, where conditions were given  for the existence of a differentiable structure on metric measure spaces, based on Lipschitz maps, which also 
have the doubling property and admit a Poincar\'{e} inequality.  Among the numerous works that clarified, elaborated and generalized \cite{Cheeger}, we could 
point out \cite{Keith, CK, Bate, CKS}. In these works the differentiable structure appears naturally as a result of the more ``primitive" assumptions stated above and
presented in \cite{Cheeger}. What is pertinent to our purposes is that \cite{Cheeger} discovered that Sobolev spaces of functions  on such metric measure 
spaces, which seem to be the most relevant from a physical viewpoint,  turn out to be reflexive. Moreover they admit a uniformly convex norm. 
We will elaborate on the first condition in this Subsection, and the second in the next Section. \\     


\noindent{\small\bf Reflexive Spaces.} \ 
Let \ $\mathcal{X}$ \ be a Banach space and let \ $B_\mathcal{X}$ \ indicate its closed unit ball, namely
\begin{equation} 
          B_\mathcal{X} \ = \ \{x\in\mathcal{X}: \ \  \|x \| \leq 1 \}
\end{equation}
The dual of $\mathcal{X}$, denoted by \ $\mathcal{X}^\prime$, \ is the space of (real or complex valued) continuous linear functionals of $\mathcal{X}$, namely 
an element of \ $\mathcal{B} (\mathcal{X}, \mathbb{R})$ or $\mathcal{B}(\mathcal{X}, \mathbb{C})$. \  Examples of such dual spaces are \ $(c_0)^\prime = l^1, \ 
(l^1)^\prime = l^\infty, \ (L^p(\mathbb{R}^n))^\prime = L^q (\mathbb{R}^n),  \ 1 < p < \infty, \ p^{-1} + q^{-1} = 1$ \ etc.    
In case $\mathcal{X}$ is finite-dimensional, the closed unit ball of its dual
$B_{\mathcal{X}^\prime}$ is the polar body of the unit ball of $\mathcal{X}$, namely
\begin{equation}    
     B_{\mathcal{X}^\prime} \ = \ B_\mathcal{X}^\circ
\end{equation}
Generalizing, one can  define the bi-dual of $\mathcal{X}$ as the dual of $\mathcal{X}^\prime$. These dualities induce a natural linear mapping 
$F: \  \mathcal{X} \rightarrow (\mathcal{X}^\prime) ^\prime$ \ given by
\begin{equation}
     F(f) \ = \ f(x)
\end{equation}
for \  $x\in\mathcal{X}$ \ and \ $f\in\mathcal{X}^\prime$ \ being its dual. 
There is no a priori reason why the double dual of $\mathcal{X}$ should be equal to $\mathcal{X}$. Usually $\mathcal{X} \ \subset (\mathcal{X}^\prime)^\prime$. 
A simple example of this inclusion is that the dual of $c_o$ is $l^1$ and the dual of $l^1$ is $l^\infty$. Therefore the inclusion $F: c_o \rightarrow l^\infty$ 
is not surjective. If, however, it happens that under the canonical map $F$
\begin{equation}
\mathcal{X} \ = \ (\mathcal{X}^\prime)^\prime 
\end{equation}
then the Banach space $\mathcal{X}$ is called reflexive. It is important that $\mathcal{X}$ is isometric to $(\mathcal{X}^\prime)^\prime $ under the canonical embedding (15).
It is possible for a non-reflexive space to be isometric to its bi-dual; an example is provided by James' space. Obviously every finite-dimensional Banach space is reflexive,
due to the rank-nullity theorem. It should be immediately noticed that reflexivity is a topological property and not a property of the norm of a particular space.
This is quite important given the fact that sometimes we will consider renormings of Banach spaces for the reasons mentioned in Subsection 2.2.\\

An example of reflexive Banach spaces are the Lebesgue spaces \  $L^p(\mathbb{R}^n), \ 1<p<\infty$. \ 
Reflexive Banach spaces have numerous desirable  properties: one can mention, for instance 
\begin{enumerate}
 \item the dual of reflexive space is reflexive.
 \item the closed subspaces of reflexive spaces are reflexive.
 \item the quotient spaces of reflexive spaces are reflexive, etc. 
\end{enumerate}
The question that comes to mind is whether there are physical reasons why a linear space should be reflexive. This is unclear in our opinion, 
beyond the desirable mathematical properties previously mentioned. It is not clear, for instance, what would be the physical consequences if the bi-dual
of a Banach space were dense, rather than surjective, under the canonical mapping (15). \\

Discrete spacetime and``internal" symmetries, such as parity, time-inversion and charge conjugation whose ``double dual" is the identity, namely idempotent operations,  
have played and continue  to play an important role in several branches of Physics. Despite this, it is not clear to us if the linear functional duality (16) has or reflects some 
deeper physical origin. One would certainly not want to preclude spaces such as $L^1(\mathbb{R}^n)$ or $L^\infty (\mathbb{R}^n)$ from being 
used as functional spaces in applications due to their non-reflexivity. The actual question that is relevant to our purposes is whether such functional spaces have 
 anything to do with the determination  of the Euclidean metric of space-time. The work of \cite{Cheeger} and subsequent developments seem to point out 
 that reflexive (Sobolev) spaces may have some special geometric significance under the assumptions of his work. 
 For this reason when combined with the implications of reflexivity for convexity and smoothness,  we will restrict our attention to reflexive Banach spaces only, 
 in the sequel. \\             


\noindent{\small\bf Finite representability.} \ 
One may be able to demand a stronger property along the lines of reflexivity, from physically relevant Banach spaces, for the purposes of determining the space-time metric,  
First a  definition: a Banach space \ $\mathcal{X}$ \ is  finitely representable in a Banach space \ $\mathcal{Z}$ \ if for every 
\  $\epsilon >0$ \ and for every finite-dimensional subspace \ $\mathcal{X}_0 \subset \mathcal{X}$ \  there is a subspace \ $\mathcal{Z}_0 \subset \mathcal{Z}$ \ 
such that \  $d_{BM} (\mathcal{X}_0, \mathcal{Z}_0) < 1+ \epsilon$. \   This essentially means that any finite-dimensional subspace of $\mathcal{X}$ can be 
represented, almost isometrically, in $\mathcal{Z}$. Equivalently, one controls the distortion of the embedding of every finite-dimensional subspace of $\mathcal{X}$ 
into $\mathcal{Z}$. From a physical viewpoint the above definition may be of interest, since it is at the confluence of two ideas:  one has to do with the fact that 
based on quantum physics, or on the statistical interpretation of theories of many degrees of freedom,  one may have to reconsider or even dispense  
with the concept of strict, ``point-wise", equality. Instead one should think much more along the lines of probabilistic equivalence, something that of course needs 
further qualifications. From such a perspective though, an approximate rather than strict, demand for isometry such as required in the definition of finite representability 
of a finite dimensional linear space is not unreasonable. The second idea relies on the fact that in any physical measurement we have a finite number of pieces of data on 
which to rely. As a result, the infinite dimensional spaces are excellent mathematical models, but from a very pragmatic perspective we  see only their finite subspaces
and then we mentally and technically extrapolate to the infinite dimensional counterparts. From this viewpoint, properties of finite dimensional vector spaces is 
the most of what someone can realistically expect to have to deal with in physical applications. \\

                       \vspace{5mm}


\noindent{\small\bf Super-reflexive spaces.} \ 
A Banach space \ $\mathcal{X}$ \ is called super-reflexive if every Banach space which is finitely representable 
in $\mathcal{X}$ is reflexive. Equivalently, a  Banach space 
$\mathcal{X}$ is super-reflexive if no non-reflexive Banach space $\mathcal{Y}$ is finitely representable in $\mathcal{X}$. Examples of  
super-reflexive spaces, pertinent to our discussion, are the Lebesgue spaces \ $L^p(\mathbb{R}^n), \ 1 < p < n$. Super-reflexive spaces are reflexive, but the converse is 
not true. Super-reflexive spaces have numerous desirable properties, from a physical viewpoint, some of which will be encountered in the next Section as they are 
pertinent to convexity and smoothness properties. One  property is that if a Banach space is isomorphic to a super-reflexive space then it is itself super-reflexive. 
Another useful property is is that a Banach space \ $\mathcal{X}$ \ is super-reflexive if and only if its dual \ $\mathcal{X}^\prime$ \ is super-reflexive.  The super-reflexivity 
of Banach spaces is a property which allows the structure of infinite dimensional Banach spaces to be determined by the embedding properties of its finite-dimensional 
subspaces. Since $c_0$ and $l^1$ are not reflexive Banach spaces, they are not super-reflexive either. For completeness, we mention  there are reflexive spaces 
are not necessarily super-reflexive: indeed consider a Banach space such that  $L^\infty (\mathbb{R}^n)$ is finitely representable in it; then it cannot be not super-reflexive.
In closing, one would like to notice that super-reflexivity, very much like reflexivity, is a topological property: as it does not really depend on the specific norm with which the  
underlying linear space is endowed.\\ 


\noindent{\small\bf Super-properties.} \ 
One can be more general at this point and talk about ``super-properties", a term that we will occasionally use in the sequel. These were defined by R.C. James in 
\cite{James-Super}.  Here, we follow the excellent ``pedestrian" exposition of \cite{Maurey}. Consider a property $P$ that is valid on a Banach space $\mathcal{X}$. 
Consider two finite-dimensional subspaces $\mathcal{Y, Z} \subset \mathcal{X}$ and numbers $n_P(\mathcal{Y}), n_P(\mathcal{Z})$ respectively such that 
\begin{equation}
        n_P(\mathcal{Y}) \ \rightarrow \  n_P(\mathcal{Z})  \hspace{10mm}   \mathrm{as} \hspace{10mm}  d_{BM} (\mathcal{Y}, \mathcal{Z})\rightarrow 1                        
\end{equation}
Probably the most straightforward case of this occurrence is when a relation like 
\begin{equation}
     n_P (\mathcal{Z}) \  \leq \ d_{BM} (\mathcal{Y}, \mathcal{Z}) \  n_P (\mathcal{Y}) 
\end{equation}
holds. Then the Banach space $\mathcal{X}$ has the property $P$ when
\begin{equation}  
    n_P(\mathcal{X}) \ = \ \sup_{\mathcal{Y}} \  n_P(\mathcal{Y}) \ < \infty
\end{equation}
where the supremum is taken over all finite dimensional subspaces of $\mathcal{X}$. \ Whether a property $P$ holds for the Banach space $\mathcal{X}$ 
evidently depends on the family $\mathfrak{F}(\mathcal{Y})$ of all finite-dimensional spaces $\mathcal{V}$ such that 
\begin{equation}  
    \forall \ \epsilon > 0 , \    \exists  \ \    \mathcal{Y} \subset \mathcal{X} : \ \ d_{BM} (\mathcal{V}, \mathcal{Y}) \ <  \ 1+ \epsilon
\end{equation}
In this terminology, finite representability of $\mathcal{Y}$ in $\mathcal{X}$ amounts to $\mathfrak{F}(\mathcal{Y}) \subset \mathfrak{F}(\mathcal{X})$. 
The property $P$ is called a super-property if whenever a Banach space $\mathcal{X}$ has $P$,  then every Banach space $\mathcal{Y}$  finitely representable in 
$\mathcal{X}$ also has $P$.\\
 
 
\noindent{\small\bf The Radon-Nikod\'{y}m property.} \ 
An additional property which is quite desirable at the technical level, and which is extensively used in Statistical 
Mechanics, where it is usually taken for granted,  
is the Radon-Nikod\'{y}m property.  It basically provides a way to make a transition between two different measures in a measure space. From a certain viewpoint, 
it can be seen as a generalization of the change of variables formula employed in multivariable calculus integration. For ``practical purposes", it states that one can use a 
function alongside the volume of a manifold as equivalent to any absolutely continuous measure. Such a density function is the micro-canonical distribution employed 
extensively in equilibrium Statistical Mechanics.  This mathematical result extends to vector-valued measures as follows:  consider a probability space ($\Omega, \mu$) 
with a $\sigma$- algebra of (Borel) sets $\Sigma$, \  $U\in \Sigma$ \  and a  Banach space \  $\mathcal{X}$. \  Let the vector-valued measure \  $\nu: \Sigma \rightarrow 
\mathcal{X}$ \  be countably additive and of bounded variation. Then there is a (Bochner) integrable function \  $f: \Omega \rightarrow \mathcal{X}$ \  such that 
\begin{equation}  
       \nu (U) \ = \  \int_U \ \ f \ d\mu
\end{equation}
For details and the generalization of this concept, from a geometric viewpoint, see \cite{CK}. 
As stated previously, this is a convenient technical theorem which allows one to use continuous functions $f$ as a form of a ``derivative" $d\nu / d\mu$. Such functions play an 
important role in Statistical Mechanics and quantum / thermal field theories as they do allow the coarse-grained distributions of interest to be treated as continuous rather than 
as discrete variables, something that many times simplifies the calculations. Moreover, it is certainly true that thermodynamic potentials such as the entropy have the 
drawback of being  coordinate-dependent for continuous distributions; the only known way around such a difficulty is the use of reference measures which invariably lead to 
Radon-Nikod\'{y}m derivatives.  Hence it may be a relief to know that reflexive Banach spaces $\mathcal{X}$ do obey the Radon-Nikod\'{y}m property and that the same is 
true for super-reflexive spaces: the obey the  super Radon-Nikod\'{y}m property.\\ 
      

                                                              \section{Convexity and smoothness.}
 
There are numerous characterizations, singling out inner product spaces among all Banach spaces. One can consult, for instance, the book \cite{Amir}
for a classical, extensive exposition and numerous results. We have also found a host of pertinent information in \cite{LindTzaf, BenLind}.\\

In this Section, we provide some information about one modulus of convexity and one modulus of smoothness and use them to single 
out Hilbert spaces  among all the Banach spaces of interest.\\


\subsection{Why convexity and smoothness?}

There is little doubt that convexity is a fundamental concept, whose origin and initial developments can be traced as far as the Greek antiquity schools of Geometry 
such as the Pythagoreans, having far-reaching consequences in many branches of Mathematics. We are interested in aspects of convexity, in this work, mainly 
in the context of vector spaces \cite{Rock} with our approach oriented toward the infinite dimensional cases.\\     

Convexity enters dynamics very early, both historically and at a stage of its development. It is present  as early as Newton's equations, at least. 
A particle trajectory in 3-dimensional space ``bends" locally  in the general direction of the total force acting on the particle. In other words, the total force acts toward 
the ``convex interior", vaguely speaking, of the curve.  The resulting acceleration is associated 
to the curvature of the trajectory in space. Such a curvature is an extrinsic concept though, namely it depends on the way the curve has been embedded in 
3-dimensional space.  Incidentally, the intrinsic geometry of a line is trivial. However this hints at a strong connection between curvature and convexity of the embedded curve. 
One can generalize this observation for higher dimensional sub-manifold embeddings. Such embeddings become crucial
in the configuration or in the phase space of a system. Consider, for instance, an isolated system; its energy os constant. 
Hence, as is well-known, its evolution takes place in a 
co-dimension one sub-manifold of its configuration space. This can be seen as a simple example of an embedding. 
The quantity characterizing such embeddings locally, in Riemannian geometry, is the second fundamental form or its closely associated shape operator.   
This topic is a classical one. For an overview and many references, see the recent thesis \cite{Verp}. The statement of interest, for our purposes, is that
if one considers a compact hyper-surface $\mathcal{M}$ of $\mathbb{R}^n$ endowed with the induced Euclidean metric, 
and the second fundamental form of this embedding is positive-definite everywhere on $\mathcal{M}$, then $\mathcal{M}$  bounds a convex 
subset of $\mathbb{R}^n$ \cite{Bishop}. This provides a local characterization of convexity for embeddings and can be seen as a generalization of 
the kinematical framework that originated with Newton's equations.\\          

One also expects convexity to be related to smoothness at least in the context of finite-dimensional vector spaces $\mathcal{V}$. A hand-waving argument that may illustrate  
this relationship is as follows: Consider a curve $\gamma: [0,1] \rightarrow \mathcal{V}$ which is locally rectifiable and which is arc-length parametrized by $s\in [0, 1]$.  
For simplicity, and in order to make this argument more clear,  assume that $\gamma$  rests on a 2-plane, at least in some neighborhood $U_x$ 
around a point $x$ corresponding to $\gamma (s_0)$. Then consider the osculating circle of $\gamma$ at $x$: this is a circle with a common tangent to the tangent 
$\frac{d\gamma (s_0)}{ds}$  whose center is along the normal line to the tangent in the $2$-plane around $U_x$. The radius of the osculating circle is equal to the 
radius of curvature $R (s_0)$ at $x$. The smaller        
the radius of curvature $R(s_0)$ the more ``steeply" the curve turns. Now assume that  $R(s_0) \rightarrow 0$. This will result in the formation of a  corner
at $\gamma (s_0)$ so $\gamma$ will no longer be differentiable at $s_0$.  At the same time, intuitively speaking,  the ``amount of convexity", a concept 
that really needs to be made precise,  of $\gamma (s_0)$ will increase as $R(s_0) \rightarrow 0$. A word of caution at this point: using Clarkson's modulus of convexity    
which is a primary object  of interest in this work, such a shrinking of the radius of curvature would leave that modulus unaffected. Still the statements regarding convexity in this 
paragraph were  meant to be heuristic and suggestive, rather than precise, in order to visually illustrate our motivation for the use of convexity and smoothness and their 
inter-relation, in the rest of this work. \\   
  
Convexity can also be seen at the level of Einstein's field equations of General Relativity. These are, in 4-dimensions,
\begin{equation}
            R_{\mu\nu} - \frac{1}{2} R g_{\mu\nu} \ = \ \frac{8\pi G}{c^4} T_{\mu\nu}
\end{equation}  
assuming that the cosmological constant \ $\Lambda = 0$ \ and \ $\mu, \nu = 0,1,2,3$. 
Consider the null energy condition \cite{CGP}, which is (arguably) the most fundamental of the energy conditions \cite{Parikh}, 
\begin{equation} 
      T_{\mu\nu} l^\mu l^\nu \ \geq \ 0
\end{equation}  
where $l$  is a null vector. (23) amounts to essentially demanding that
\begin{equation}
     R_{\mu\nu} l^\mu l^\nu \ \geq \ 0
\end{equation}
(24) can be interpreted as a (mean) convexity condition in a null direction: one way is to see that the Ricci tensor involves two derivatives of the components of the metric 
tensor hence the non-negativity of (24) signifies convexity in analogy with the case of functions. Such convexity can also be seen in a Riemannian context. 
There, the non-negativity of the Ricci curvature (24) can be shown \cite{CEMcS} to imply a generalized Brunn-Minkowski inequality, which is essentially a statement 
about the concavity, the convexity of the opposite function, of the volume in Euclidean and Riemannian spaces.  \\


\subsection{A modulus of convexity.}

\noindent{\small\bf Convex sets.} \ 
To be more precise and attempt to make the exposition somewhat self-contained, we present the following well-known definitions and statements. 
Consider $\mathcal{V}$ to be a vector space over $\mathbb{R}$ or $\mathbb{C}$. A subset 
$\mathcal{A} \subset \mathcal{V}$ is called (affinely) convex if 
\begin{equation}  
       \{ ta_1 + (1-t)a_2, \ \ t \in [0,1]  \} \  \subset \  \mathcal{A}, \ \  \ \ \ \  \forall \ a_1, a_2 \in \mathcal{A}
\end{equation}
Equivalently, for every \ $n \in \mathbb{N}$ \ and for every \ $t_0, t_1, \ldots, t_n \in [0, 1]$ \ such that \ $t_0 + t_1 + \ldots + t_n = 1$ \
and for every \ $a_0, a_1, \ \ldots,  a_n \in \mathcal{A}$, \ we have \ $t_0a_0 + t_1a_1 + \ldots + t_na_n \in \mathcal{A}$. \ Obviously \ $\mathbb{R}^n$ \ 
as well as its linear and affine subspaces are convex. The same conclusion holds about  open and closed unit balls in normed vector spaces. \\ 


\noindent{\small\bf Convex functions.} \ Convex functions can be considered as generalizations of convex sets. Let $\mathcal{A} \subset \mathcal{V}$ 
be a convex subset of the linear space $\mathcal{V}$
and let $f: \mathcal{A} \rightarrow \mathbb{R}$ be a function. The epigraph of $f$ is defined to be the set
\begin{equation}    
     Ep(f) \ = \  \{ (a, t) \in \mathcal{A} \times \mathbb{R}:  \ \ f(a) \leq t \}
\end{equation}
Then, such a function $f$ is called convex, if \ $Ep(f)$  \  is a convex subset of \ $\mathcal{A} \times \mathbb{R}$. \ Equivalently, if for all \ $a, b \in \mathcal{A}$ \  
 and  \ $t \in [0,1]$, \ such \ $f$ \ satisfies the inequality
\begin{equation}
     f(ta+(1-t)b) \   \leq \  (1-t) f(a) + t f(b)
\end{equation}  
The combination of homegeneity and the triangle inequality shows immediately that a norm $\| \cdot \|$ on $\mathcal{V}$ is a convex function. Other convex functions 
are the distance functions between two lines in Euclidean space ($\mathbb{R}^n$ endowed with the Euclidean metric) and, more generally, distance functions on  
metric spaces of negative curvature. Moreover, if in a normed (more generally: a metric) space \ $\mathcal{V}$ \  with \ $\mathcal{A} \subset \mathcal{V}$  \  and for  
all \ $a\in\mathcal{V}$, \ one defines 
\begin{equation}     
       d_\mathcal{A} (a) \ = \ \inf_{b\in \mathcal{A}} \|a-b\|
\end{equation}
when \ $\mathcal{A}$ \ is a nonempty closed convex subset of \ $\mathcal{V}$, \  then the distance function \ $d_\mathcal{A}: \mathcal{V} \rightarrow \mathbb{R}_+$ \  is 
convex. \\

Convex functions have many nice properties: they are semicontinuous and  almost everywhere differentiable, they possess left and right derivatives (which however need 
not coincide), limits of sequences of convex functions defined on convex sets are convex functions, they have a unique minimum and they obey the local-to-global property, 
namely a locally convex function is actually (globally) convex.  It is properties like these that make convex functions so useful and widespread in Physics and, in particular, 
in Analytical Mechanics and Thermodynamics.\\


\noindent{\small\bf A modulus of convexity.} \  
A next logical  step is to find a way to quantify the extent of convexity of a set or of a function. Naturally, determining such a ``modulus of convexity" is not a unique process 
and it involves making certain choices. Simplicity and computability in, at least, simple cases are usually good guidelines, as far as physical applications are concerned. 
A relatively recent list of such moduli which is quite extensive, even if  not necessarily comprehensive, can be found in \cite{Fuster}. \\

The oldest and most studied among the moduli of convexity is due to J.A. Clarkson \cite{Clarkson, Day}. To define it, consider the normed linear space 
\ ($\mathcal{X}, \| \cdot \|$), \  let \ $B_\mathcal{X}$ \  indicate its closed unit ball, as before, and let  \  $\varepsilon \in [0,2]$. \  
The modulus of local convexity is defined by 
\begin{equation}
          \delta (x, \varepsilon) \ = \ \inf \left\{ 1 - \left\| \frac{x+y}{2} \right\|: \ y\in B_\mathcal{X}, \  \| x-y \| \geq \varepsilon \right\}   
\end{equation}
Then Clarkson's modulus of convexity of \ $\mathcal{X}$ \  denoted by \  $\delta_\mathcal{X}: [0,2] \rightarrow [0,1]$ \  is defined as     
\begin{equation}
      \delta_\mathcal{X} (\varepsilon) \ = \ \inf \{ \delta (x, \varepsilon ): \ x\in B_\mathcal{X} \}
\end{equation}
This can be equivalently expressed  as 
\begin{equation} 
     \delta_\mathcal{X} (\varepsilon) \ = \  \inf \left\{ 1 - \left\| \frac{x+y}{2} \right\| : \  x,y\in\mathcal{X}, \  \| x \| = \| y \| = 1, \ \|x-y \| = \varepsilon     \right\}  
\end{equation}
It should be immediately noticed that this modulus of convexity is really a property of the 2-plane spanned by $x, y \in \mathcal{X}$ which is then inherited by 
$\mathcal{X}$. This should look familiar: the curvature of an $n$-dimensional Riemannian manifold $\mathcal{M}$  is a genuinely 2-dimensional concept 
which is actually formulated on its Grassmann manifold $G_{2,n}(\mathcal{M})$. The (sectional) curvature expresses the deviation of the metric of $\mathcal{M}$ from its 
flat counterpart \cite{Gr-GMC}.  The characteristic of convexity of \  ($\mathcal{X}, \| \cdot \|$)   \  is defined as \cite{Goebel}  
\begin{equation}
     \varepsilon_0 (\mathcal{X} ) \ = \  \sup \{ \varepsilon \in [0, 2]: \  \delta_\mathcal{X} (\varepsilon ) = 0 \} 
\end{equation}


\noindent{\small\bf Uniformly convex spaces.} \ 
A Banach space \ ($\mathcal{X}, \| \cdot \|$) \ is uniformly convex when it has non-zero modulus of convexity, namely \  $\delta_\mathcal{X} (\varepsilon) > 0, \ \ 
\varepsilon \in (0, 2]$ \ or equivalently when \  $\varepsilon_0 (\mathcal{X}) = 0$. \ Geometrically, the idea of the definition is simple: uniformly convex spaces 
have a unit ball \  $B_\mathcal{X}$ \ whose boundary unit sphere $S_\mathcal{X}$ does not contain any (affine) line segments. 
Roughly speaking: the further away from containing an affine segment  \ $S_\mathcal{X}$ \  is, the higher the modulus of convexity of \  $\mathcal{X}$  \  is. 
It should be noticed that according the  D.P. Milman \cite{Milman} - B.J. Pettis \cite {Pettis} theorem,  uniformly convex spaces are reflexive. 
Actually one can see that uniformly convex spaces  are actually super-reflexive. Hence, if one deems reflexivity or super-reflexivity
to be a desirable, or pertinent, property in an argument about the Euclidean nature of the space-time metric, as was stated above, 
then confining their attention to uniformly convex Banach spaces will not miss this property.\\

From a different viewpoint, it is known from the work of P. Enflo \cite{Enflo} and R.C. James \cite{RCJames} that $\mathcal{X}$ is super-reflexive if and only if it has an 
equivalent uniformly convex norm. ``Equivalence" in this theorem is meant to be understood in the sense described in Subsection 2.2. Therefore, insisting on having 
super-reflexivity of the underlying functional spaces that determine/induce the space-time metric usually has to allow for a change of their norm to a uniformly convex one, 
if this is feasible. This change may have substantial physical implications for the underlying model. If however, as mentioned above, 
one is only interested in large-scale ``coarse" phenomena that ignore spatially ``small" details and arise as a result of statistical averaging of many degrees of freedom, 
then one believes that such a renorming will leave the macroscopic quantities of interest unaffected. \\ 


\noindent{\small\bf Modulus of convexity of Lebesgue spaces.} \  
Explicitly calculating the modulus of convexity for specific Banach spaces has proved to be more difficult than one might have naively anticipated. 
This is one reason why so many different moduli of convexity have been defined over the decades, after Clarkson's work \cite{Fuster}.
For completeness, we mention that for \ $p=1$ \ and for \ $p = \infty$, \  $\delta_{L^p} = 0$ \ as these two Banach spaces are not uniformly convex. 
The fact that the other Lebesgue spaces \ $L^p(\mathbb{R}^n), \ 1 < p < \infty$ \ are uniformly convex was already known to \cite{Clarkson}. For a simpler and more 
recent proof see \cite{HO}. However, the explicit asymptotic form of their modulus of convexity was  determined by \cite{Hanner} who relied on the inequalities 
bearing his name \cite{Hanner, BKL} to reach his result. Hanner proved that 
\begin{equation}         
   \delta_{L^p} (\varepsilon) \ = \ \left\{                            
                                 \begin{array}{ll}
                                 (p-1) \frac{\varepsilon^2}{8} + o(\varepsilon^2), & \mathrm{if} \ \ 1 < p \leq 2\\ 
                                                              &   \\
                                 \frac{1}{p} \left( \frac{\varepsilon}{2} \right) ^p + o(\varepsilon^p), & \mathrm{if} \ \  2 \leq p < \infty 
                                 \end{array}
                                           \right.
\end{equation}
One cannot fail to notice the ``phase transition" in the asymptotic behavior of this  modulus of convexity occurring around \ $p=2$, \  namely around the Hilbert space case.
An inevitable question is  how generic such a behavior might be, among all Banach spaces. If it is, then it would be an initial indication that Hilbert spaces may be ``special",
from a convexity viewpoint. It turns out that this is essentially true, and this is the statement on which the main argument of the present work relies.  
This is the realization that among all Banach spaces, the Hilbert space is the ``most" convex. 
To be more precise, it was proved in \cite{Nord} that among all Banach spaces \ $\mathcal{X}$, \ 
the Hilbert spaces \ $\mathcal{H}$ \ have the largest modulus of convexity,  namely
\begin{equation}
     \delta_\mathcal{X} (\varepsilon ) \ \leq \ \delta_\mathcal{H} (\varepsilon)
\end{equation}
where 
\begin{equation}
       \delta_\mathcal{H} (\varepsilon) \ = \ \frac{2 - \sqrt{4 - \varepsilon^2}}{2} 
\end{equation}
It is also interesting to notice that the results of \cite{Nord} when combined with those of \cite{Day2, Sch} prove that if a linear  space \ $\mathcal{X}$ \  is such that 
the equality in (34) holds, then \  $\mathcal{X}$ \ is an inner product space. \\


\noindent{\small\bf Modulus of convexity and equivalences.} \ 
It should be noted that the modulus of convexity \ $\delta_\mathcal{X} (\varepsilon)$ \ is not necessarily itself a convex function of $\varepsilon$. 
What we know, for instance, is that  for an infinite dimensional uniformly convex Banach space $\mathcal{X}$, we have that 
\begin{equation}
           \delta_\mathcal{X} (\varepsilon) \ \leq \ C \varepsilon^2  
\end{equation}
For a Hilbert space we have (35) which is, obviously, compatible with (36). So, in a quantitative sense, the closer \ $\delta_\mathcal{X}$ \ is to \  $\varepsilon^2$ \ 
for a normed space \ $\mathcal{X}$, \  the closer \ $\mathcal{X}$ \ is being maximally convex, hence the closer it is to being a Hilbert space \ $\mathcal{H}$. \ 
T. Figiel \cite{Figiel} showed that one  can consider as a modulus of convexity instead of \  $\delta_\mathcal{X}(\varepsilon)$  \ the greatest convex function \  
$\tilde{\delta}_\mathcal{X} (\varepsilon)$, \ namely
\begin{equation}  
       \tilde{\delta}_\mathcal{X} (\varepsilon) \ \leq \ \delta_\mathcal{X} (\varepsilon)  
\end{equation}
Then \cite{Figiel}
\begin{equation}
   c_1 \delta_\mathcal{X} (c_2\varepsilon) \ \leq  \ \tilde{\delta} (\varepsilon)
\end{equation}
for constants \ $c_1 > 0 , \  c_2 > 0$. \\

 The re-definition employed in \cite{Figiel} is reminiscent of the equivalence of growth functions in geometric group theory
when one has to decide whether such a growth function is exponential, polynomial or has an intermediate growth rate \cite{Grig}. The reason for such similarity is quite 
clear: in geometric group theory, as in our case, one really cares about equivalences that may distort distances, but leave large-scale details of the structure invariant. 
The metric equivalence employed there is that of quasi-isometric maps, which distorts distances between two metric spaces \ $(\mathfrak{X}_1, d_1), \ 
(\mathfrak{X}_2, d_2)$ \    according to 
 \begin{equation}
        \frac{1}{c_1} d_1(x,y) - c_2    \   \leq \  d_2(f(x), f(y))  \  \leq   \ c_1 d_1(x,y) + c_2,  \hspace{10mm} \forall \ x,y \in \mathfrak{X}_1
 \end{equation}    
where \ $c_1 > 1$ \ and \ $c_2 > 0$. \ The difference between (6) and (39) is the presence of $c_2$ in (39) which completely ignores structures whose length 
is smaller than $c_2$.
We have previously employed such equivalences in our work in \cite{NK1, NK2} in our attempt to determine the dynamical basis of a power-law entropy, something that 
may have implications for the derivation of the metric of space-time from microscopic models of quantum gravity. The use of such maps acquires even greater significance
when one considers that it is intimately related to properties of (Gromov) hyperbolic spaces. After a $3+1$ decomposition, 3-dimensional manifolds represent space-like 
hyper-surfaces in spacetimes. Following the results of the Thurston geometrization program (see \cite{Scott} for an overview), it seems that ``most" of the 3-dimensional 
manifolds are hyperbolic. Hence quasi-isometries and similar ideas may be relegated to a central role in determining classical/long-distance structures of the (4-dimensional) 
spacetimes from their microscopic/quantum foundations.  \\   


\subsection{A modulus of smoothness and a duality.}

A quantity intimately related to convexity, in the context of normed linear spaces, is smoothness. In line with convexity, one also needs a way to quantify the 
amount of smoothness of a linear space. Once more, there is no unique way of how to go about constructing such a modulus of smoothness and actually 
several such moduli of smoothness have been constructed over the years \cite{Fuster}.\\


\noindent{\small\bf A modulus of smoothness.} \ 
The oldest and most studied modulus of smoothness is due to M.M. Day \cite{Day} and J. Lindenstrauss \cite{Linden}.  The modulus of smoothness of a normed space
\  ($\mathcal{X}, \| \cdot \|$) \  is a function \  $\rho_\mathcal{X}: [0, \infty) \rightarrow \mathbb{R}$ \ which is  defined by
\begin{equation}     
       \rho_\mathcal{X} (t) \ = \  \sup \left\{ \frac{\|x+ty\| + \| x-ty\|}{2} - 1, \ \  \ \ x, y \in B_\mathcal{X} \right\} 
\end{equation}
This can be alternatively defined as
\begin{equation}
   \rho_\mathcal{X} (t) \ = \ \sup \left\{ \frac{\|x+y\| + \|x-y \|}{2} - 1 :  \ \ x\in B_\mathcal{X}, \  \| y \| \ \leq t  \  \right\} 
\end{equation}
or as
\begin{equation}
   \rho_\mathcal{X} (t) \ = \ \sup \left\{ \frac{\|x+y\| + \|x-y \|}{2} - 1, \ \ \ \ x,y \in S_\mathcal{X}  \right\}  
\end{equation}
One defines the coefficient of smoothness of \ $\mathcal{X}$ \ by 
\begin{equation}
       \rho_0 (\mathcal{X}) \ =  \ \lim_{t\rightarrow 0^+} \frac{\rho_\mathcal{X}(t)}{t} 
\end{equation}


\noindent{\small\bf Uniformly smooth spaces.} \ 
The Banach space \  ($\mathcal{X}, \| \cdot \|$) \ is called uniformly smooth if \ $\rho_0 (\mathcal{X}) = 0$. \ One sees immediately that uniform smoothness is a 
point-wise property and is essentially  2-dimensional, as is the case for uniform convexity. In a pictorial sense, this modulus of smoothness captures the fact that the unit 
sphere \ $S_\mathcal{X}$ \ of the Banach space \ $\mathcal{X}$ \  is smooth, i.e. that it has not corners. More issue on this issue is discussed below.\\

Before continuing it may be worth developing a pictorial idea about uniform convexity and uniform smoothness. Consider, for instance, a square with rounded corners.
This is not uniformly convex, since it includes line segments in its boundary, but it lacks corners, so it is uniformly smooth. Consider now the, assumed non-empty and 
not a point, intersection of two equal radius disks: it is uniformly convex as its boundary contains no line segments, but it is not uniformly smooth since at the two 
intersection points of the two circles which are the boundaries of the two disks, the figure has corners.\\     


\noindent{\small\bf Modulus of smoothness of Lebesgue spaces.} \ 
Unlike the modulus of convexity, the modulus of smoothness is a convex function, essentially by definition. The asymptotic behavior of the modulus of smoothness 
of the Lebesgue spaces \ $L^p(\mathbb{R}^n)$ \ were also calculated in \cite{Hanner}. O. Hanner found that 
\begin{equation} 
     \rho_{L^p} (t) \ = \ \left\{   
                             \begin{array}{ll}
                                      \frac{t^p}{p} + o(t^p), & \ \ \mathrm{if}  \ \ 1 <  p  \leq 2 \\
                                                                        &         \\
                                      \frac{p-1}{2} \ t^2 + o(t^2), & \ \  \mathrm{if} \  \  2 \leq  p < \infty                                              
                              \end{array}
                                    \right.
\end{equation}
The same comments and thoughts regarding the phase transition around \ $p=2$  \ apply here as in the case of the modulus of convexity as stated above.
The counterpart of (34) was also established, and it states that that for any Banach space \ $\mathcal{X}$ \ one has 
\begin{equation}  
       \rho_\mathcal{X} (t) \  \geq \ \rho_\mathcal{H}  (t)
\end{equation}  
Therefore Hilbert spaces are the least smooth among all Banach spaces. Moreover, the exact same statement as the one following (35) 
applies for the modulus of smoothness: if a Banach space $\mathcal{X}$ obeys
\begin{equation}
      \rho_\mathcal{X} (t)  \ = \ \rho_\mathcal{H} (t)
\end{equation}
then $\mathcal{X}$ is an inner product space. The modulus of smoothness, very much like for the modulus of convexity, of a Hilbert space is a result of the validity of the 
parallelogram equality and it is given by
\begin{equation}
    \rho_\mathcal{H} (t) \ = \ \sqrt{1+t^2} - 1 
\end{equation}
The analogue of (36) is that for an infinite dimensional Banach space $\mathcal{X}$ one has 
\begin{equation}    
       \rho_\mathcal{X} (t) \ \geq \ c t^2
\end{equation}  

The Milman-Pettis theorem established, in addition to  uniform convexity, that uniformly smooth spaces are reflexive. Even though the dual of a super-property may not be 
a super-property, using the ``convexity-smoothness" duality of the next paragraph, one sees that uniformly smooth spaces are indeed super-reflexive. 
Analogous  things  can be stated about the uniform smoothability of a Banach space as were stated about their uniform convexifiability, with equivalent norms.  \\


\noindent{\small\bf A duality and the Legendre-Fenchel transforms.} \ 
One observes form the above that Clarkson's modulus of convexity and the Day-Lindenstrauss modulus of smoothness appear to behave very much like dual concepts.
The fact is that this suspected duality is true. More precisely,  \cite{Linden} proved that a Banach space \ $\mathcal{X}$ \  is uniformly convex if and only if its dual 
\  $\mathcal{X}^\prime$ \  is uniformly smooth. The exact relation between the corresponding moduli is given by the Legendre-Fenchel transform  
\begin{equation} 
      \rho_{\mathcal{X}^\prime} (t) \  = \  \sup \left\{ \frac{\varepsilon t}{2} - \delta_\mathcal{X} (\varepsilon), \ \ 0 \leq \varepsilon \leq 2 \right\}
\end{equation}
or, equivalently, 
\begin{equation}
   \rho_\mathcal{X} (t) \ = \ \sup \left\{ \frac{\varepsilon t}{2} - \delta_{\mathcal{X}^\prime} (\varepsilon), \ \ 0 \leq \varepsilon \leq 2 \right\}
\end{equation}
Therefore, every theorem valid for convexity has a dual analogue valid for smoothness. We see, for instance, that 
\begin{equation}
       \rho_0(\mathcal{X}^\prime) \ = \ \frac{1}{2} \varepsilon_0 (\mathcal{X}),  \hspace{10mm} \rho_0 (\mathcal{X}) \ = \ \frac{1}{2} \varepsilon_0 (\mathcal{X}^\prime)
\end{equation}
which shows that a Banach space $\mathcal{X}$ is uniformly convex if and only if its dual $\mathcal{X}^\prime$ is uniformly smooth. Continuing along the lines of the 
quantification of convexity and smoothness, one can state the following: consider the Banach space $\mathcal{X}$. It turns out that 
\begin{equation}
     \delta_\mathcal{X} (\varepsilon) \ \geq \  C \varepsilon^q, \hspace{10mm} q \ \geq \ 2 
\end{equation}
in which case $\mathcal{X}$ is called a $q-$convex space.  According to a theorem of Figiel and Assouad, this is equivalent to the existence of some constant $\alpha$
such that
\begin{equation}
      \frac{1}{2}  \left( \|x+y\|^q + \|x-y \|^q \right) \ \geq \ \|x \|^q +\alpha \|y \|^q,  \hspace{10mm} \forall \ x,y \in \mathcal{X}   
\end{equation}
For smoothness, the corresponding statement is that for the Banach space $\mathcal{Y}$ one has   
\begin{equation}
      \rho_\mathcal{Y} (t) \ \leq \  c t^p,           \hspace{10mm}    1 < p \leq 2
\end{equation}
in which case $Y$ is called $p-$smooth. Then, according to a theorem of Pisier and Assouad, this is equivalent to the existence of some constant $\beta$ such that 
\begin{equation}
      \frac{1}{2} \left( \|x+y\|^p + \|x-y\|^p \right) \ \leq \ \|x \|^p + \beta \| y \|^p, \hspace{10mm} \forall  \ x,y \in \mathcal{Y}
\end{equation}
Based on the Legendre-Fenchel duality noted above, one can also state that a Banach space $\mathcal{X}$ is $q$-convex if and only if its dual $\mathcal{X}^\prime$
is $p$-smooth, where $q$ qnd $p$ are harmonic conjugates, namely
\begin{equation}
    \frac{1}{p} + \frac{1}{q} \ =   \  1 
\end{equation}


\subsection{Smoothness, derivatives and equivalences.}

The word ``smoothness" is associated, in a typical Physicist's mind, with the concept of the derivative. It may be of interest to know that the same can be stated for
the cases of the Banach spaces of our interest. Using infinite dimensional spaces though introduces additional complexities and also possible counter-intuitive 
features that should be carefully accounted for. For one, there are several possible definitions for the (directional) derivative in a normed space. 
We only need the definition of the Frech\'{e}t derivative in this work \cite{BenLind}. 
Let \  $\mathcal{X}, \mathcal{Y}$ \  be Banach spaces and let \ $f:\mathcal{X} \rightarrow \mathcal{Y}$. \ Then \ $f$ \  is called Frech\'{e}t
differentiable at \ $x\in\mathcal{X}$, \ if there is a bounded linear operator \ $A_x : \mathcal{X} \rightarrow \mathcal{Y}$ \ such that the following limit exists uniformly 
for \  $y \in S_\mathcal{X}$    
\begin{equation}
     A_x y \ = \ \lim_{t\rightarrow 0} \  \frac{ f(x+ty) - f(x)}{t}
\end{equation}
If this limit indeed exists, then it is called the Frech\'{e}t derivative of \ $f$ \ at \ $x$ \ and is indicated by \ $D_f(x)$. \ The pertinent statement is that if \ $\mathcal{X}$ \ 
is a uniformly smooth Banach space then its norm \  $f(x) =  \| x \|$ \ is Frech\'{e}t differentiable for every \ $x\in \mathcal{X} \backslash  \{ 0 \}$. In such a case, one can 
see that the linear approximation to $f$ at $x$ through $D_f(x)$ is valid, namely we can write
\begin{equation}  
   f(x+y) \ = \ f(x) + D_f(x)y + o(\| y \|)
\end{equation}
If a function is Frech\'{e}t differentiable at a point, it turns out that it is also continuous at that point, which is the Banach space analogue of a well-known and 
frequently used result of elementary calculus. \\

In addition to  the above, it turns out \cite{BenLind} that for a Banach space \ $\mathcal{X}$ \ the following statements are equivalent:
\begin{itemize}
    \item $\mathcal{X}$ is super-reflexive.
    \item $\mathcal{X}$ admits an equivalent, uniformly convex norm, whose modulus of convexity satisfies, for some \  $q\geq 2$, 
                      \begin{equation}           
                                 \delta_\mathcal{X} (\varepsilon) \ \geq  \ c_1 \varepsilon^q  
                      \end{equation}
    \item $\mathcal{X}$ admits an equivalent, uniformly smooth norm, whose modulus of smoothness satisfies, for some \  $1 < p \leq 2$ 
                      \begin{equation}
                                \rho_\mathcal{X} (t) \ \leq \ c_2 t^p
                      \end{equation}
\end{itemize}
Moreover, Asplund \cite{Asp} showed that a space which admits two equivalent norms one of which is uniformly convex and the other uniformly smooth, 
admits a third one which is equivalent to the previous two and which has both properties. As a special case, one can state that a super-reflexive space 
admits an equivalent norm which is both uniformly convex and uniformly smooth.\\


\subsection{Type, co-type and moduli.}

\noindent{\small\bf Rademacher functions.} \ 
The powers \ $q$ \ and \ $p$ \  in the lower bound of the modulus of convexity (52) and in the upper bound in the modulus of smoothness (54), respectively, 
have a nice geometric-probabilistic interpretation. To formulate it, we need to use the Rademacher functions which are defined as 
$r_i : [0,1] \rightarrow \pm 1$ \  by 
\begin{equation}
      r_i (t) \ = \ \mathrm{sign} [\sin(2^i\pi t )], \hspace{10mm} i\in \mathbb{N}
\end{equation}
We see that the Rademacher functions can be interpreted as a sequence of identically distributed random variables on  \ $[0,1]$ \ endowed with its Lebesgue 
measure, taking values \ $\pm 1$, \  each with probability \ $0.5$. \ It may be worth noting that for vectors \ $x_i, \ i=1, \ldots n$ \ in a Banach (or more generally in a 
normed) space \  ($\mathcal{X}, \| \cdot \|$), \  one has 
\begin{equation} 
       \int_0^1 \left\| \sum_{i=1}^n r_i(t)x_i \right\|^p \ dt \ \ =  \ \ \mathbb{E} \left( \left\| \sum_{i=1}^n \epsilon_i x_i \right \|^p \right)
\end{equation}
where the expectation value \ $\mathbb{E}$ \  is taken over \ $\epsilon_i = \pm 1$. \\


\noindent{\small\bf Type.} \ 
 Suppose now that for any finite number $n$ and any choice of vectors \ $x_i, \ i=1,\ldots,n$ of  $\mathcal{X}$ \ there is a constant \ $C_p > 0$ \ such that 
\begin{equation} 
       \left( \int_0^1 \left\| \sum_{i=1}^n r_i(t) x_i \right\|^p dt \right)^\frac{1}{p}  \  \leq \ C_p \left( \sum_{i=1}^n \| x_i \|^p  \right)^\frac{1}{p}  
\end{equation}   
then the Banach space \ $\mathcal{X}$ \ is said to have type \ $p$. \  The best constant \ $C_p(\mathcal{X})$ \  in the definition (63) is called type-$p$ constant of \ 
$\mathcal{X}$. \  Consider A.I. Khintchine's inequality for \ $a_i,  \   \  a_i \in \mathbb{R}, \ \ i=1, \ldots, n$, \ $0 < p < \infty$, the Rademacher functions $r_i(t)$,  and constants 
\ $A_p, \ B_p$, \ namely
\begin{equation}
   A_p \left( \int_0^1 \bigg{|} \sum_{i=1}^n a_i r_i (t) \bigg{|}^p dt \right)^\frac{1}{p} \ \leq \  \left( \int_0^1 \bigg{|} \sum_{i=1}^n a_i r_i(t) \bigg{|}^2  dt  \right)^\frac{1}{2} \ \leq \ 
           B_p \left( \int_0^1 \bigg{|} \sum_{i=1}^n a_i r_i (t) \bigg{|}^p dt \right)^\frac{1}{p}
 \end{equation}
If we assume that all vectors are equal in  (63), then by using  (64) we see that
that   \ $p\leq 2$. \ By using the triangle inequality, one can also see that \ $p\geq 1$. \ Therefore the only allowed values for a type \  
$p$ \ Banach space are \ $1 \leq p \leq 2$.\  This is equivalent to stating that the elements of the family $\mathfrak{F}(\mathcal{Z})$ 
of finite-dimensional subspaces  $\mathcal{Z} \subset \mathcal{X}$ satisfy
\begin{equation}
      \sup_{\mathfrak{F}(\mathcal{Z})} C_p (\mathcal{Z}) \   <   \ \infty
\end{equation}
which shows that \ $\mathcal{X}$ \  having type \ $p$ \ is a super-property. 
One can also see that if \ $p^\prime < p$ \ then type \ $p$ \ implies type \ $p^\prime$. \\ 


\noindent{\small\bf Co-type.} \ 
With similar notation as for type, a Banach space \ $\mathcal{Y}$ \ has co-type \ $q$ \ if there is a constant \ $C^\prime _q >0$ such that     
\begin{equation}
      \left( \int_0^1 \left\| \sum_{i=1}^n r_i(t) x_i \right\|^q dt  \right)^\frac{1}{q}  \    \geq    \  C^\prime _q \left( \sum_{i=1}^n \| x_i \|^q  \right)^\frac{1}{q}
\end{equation}
for any \ $n\in\mathbb{N}$ \ and for any set of vectors \ $x_i \in\mathcal{Y}$. \ Again, the best constant \ $C^\prime _q(\mathcal{Y})$ \ in (66) is called the co-type \ $q$ \ 
constant for \ $\mathcal{Y}$. \ This again is equivalent to stating that the elements of the family \ $\mathfrak{F}(\mathcal{W})$, \ for all finite-dimensional subspaces \  
$\mathcal{W} \subset \mathcal{Y}$ \  satisfy 
 \begin{equation}
      \sup_{\mathfrak{F}(\mathcal{W})} C_q (\mathcal{W}) \ < \infty 
 \end{equation}  
which also shows that \ $\mathcal{Y}$ \  having co-type \ $q$ \ is a super-property. By using Khintchine's inequality again, one can see that the co-type of any Banach 
space that has one has to be $q \geq 2$. \ One can also see that if \ $q^\prime > q$, \  then co-type $q$ implies co-type $q^\prime$. \\


\noindent{\small\bf Type and co-type properties.} \ 
As an explicit example of  type and co-type  we know \cite{LindTzaf} that the Lebesgue spaces $L^p (\mathbb{R}^n)$ have 
\begin{itemize}   
       \item Type \ $p$ \  and co-type \ $2$, \  if \ $1\leq p \leq 2$
       \item Type \ $2$ \ and  co-type \ $p$, \  if \ $2\leq p < \infty$  
\end{itemize}
One again cannot fail to notice the ``phase transition" in the behavior of type and co-type of \ $L^p(\mathbb{R}^n)$ \ taking place around \ $p=2$. \ 
Hilbert spaces are singled out in a stronger sense: indeed, using the parallelogram equality (38) below, one can see that a Hilbert space has type $2$ and co-type $2$. 
That the converse is also true, is a non-trivial result due to S. Kwapie\'{n} \cite{Kwapien}. \\

The type and co-type of a Banach space $\mathcal{X}$ can be seen in a variety of ways: one way is to state that they are a way of quantifying how 
far \ $\mathcal{X}$ \  is  from being a Hilbert space. The comparison of the definitions of type and co-type with the parallelogram equality that only Hilbert spaces satisfy 
\begin{equation}
    \| x+y \|^2 + \| x-y \|^2 = 2(\| x\|^2 + \| y \|^2)  
\end{equation}
is quite suggestive. In view of the last paragraph of this Subsection (see below) the same can be stated by comparing (68) to (53) and (55) 
which alternatively quantify the moduli of convexity and smoothness of \ $\mathcal{X}$. \\
    
Now, one can inquire about the robustness of the concepts of type and co-type.  
J.P. Kahane's inequality \cite{Kahane} is a vector-valued extension of A.I. Khintchine's inequality (64) and states, with the above notation, 
that for every \ $0 < p < q < \infty$ \  there is a constant \ $C(p,q) > 0$ \  such that 
\begin{equation}
      \left( \int_0^1 \left\| \sum_{i=1}^n r_i(t) x_i  \right\|^p dt \right)^\frac{1}{p}   \   \leq \  \left( \int_0^1 \left\| \sum_{i=1}^n r_i(t) x_i  \right\|^q dt  \right)^\frac{1}{q}
                                      \ \leq \  C(p,q) \left( \int_0^1 \left\| \sum_{i=1}^n r_i(t) x_i  \right\|^p dt \right)^\frac{1}{p}  
 \end{equation}
This shows that the type $p$ and co-type $q$ are properties that are maintained under an equivalent norm. Moreover, instead of using Rademacher functions
in the definitions, something which is technically convenient, one could use centered Bernoulli random variable, Gaussian random variables etc. with just a 
change in the values of the constants in the definitions \cite{MS}.  \\  


\noindent{\small\bf Type, co-type and moduli.} \ 
The relation of the type and co-type of a Banach space with its moduli of convexity and smoothness is contained in the following theorem due to T. Figiel, G. Pisier 
\cite{FiPi}: Let \ $\mathcal{X}$ \ be a uniformly convex Banach space with modulus of convexity satisfying \ $\delta_\mathcal{X} (\varepsilon) \geq C\varepsilon^q$  \ 
for some \ $q \geq 2$. \ Then \ $\mathcal{X}$ \  has co-type \ $q$. \  Let \ $\mathcal{Y}$ \  be a uniformly smooth Banach space whose modulus of smoothness satisfies 
\ $\rho_\mathcal{Y}(t) \leq c t^p$ \ for some \ $1 < p \leq 2$. \ Then \ $\mathcal{Y}$ \ has type \ $p$. \  Therefore the moduli of smoothness and convexity of a Banach space 
are bounded by the type and co-type of that Banach space, assuming that the latter exist. It may be worth noticing at this point the behavior of type and co-type 
under duality: It is known that when a Banach space \ $\mathcal{X}$ \ has type \ $p$, \  then its dual \ $\mathcal{X}^\prime$ \ has co-type \ $q$ \ where \ $p$ \ and \ $q$ \ 
are harmonic conjugates of each other. However the converse is not true without one additional  assumption. The accurate statement is that if a Banach space \ 
$\mathcal{X}$ \ has non-trivial type, and co-type \ $q$, \  then its dual \ $\mathcal{X}^\prime$ \ has type \ $p$, \  where \ $p$ \ and \ $q$ \ are harmonic conjugates 
of each other. For excellent expositions of the type and co-type of normed spaces, including proofs of all of the above statements, one may 
consult \cite{LindTzaf, BenLind, MS}.  \\    


\section{The space-time metric  from variational principles.}

One cannot fail to notice from the content of the previous Sections, the unique role that inner product (Hilbert) spaces \ $\mathcal{H}$ \  play among all Banach spaces 
\ $\mathcal{X}$ \ and, in particular, the distinct role of \ $L^2(\mathbb{R}^n)$ \ among all \ $L^p(\mathbb{R}^n)$. \  
Such Hilbert spaces are, at the same time, the most convex and least smooth among all Banach spaces. They are the only Lebesgue spaces that have the same 
type and co-type $2$. In addition, they are super-reflexive. Moreover such Hilbert spaces are the only 
self-dual Lebesgue spaces under harmonic conjugation, a basic fact reflecting properties of the polarity of convex bodies and 
of the Legendre-Fenchel transformations \cite{ArtMil}. \\


\noindent{\small\bf Functional integrals and variational principles.} \ 
 Using extremal (more accurately: stationary) properties of functionals under infinitesimal variations subject to appropriate boundary conditions has been a 
fundamental aspect of Classical and Quantum Physics since the time of Maupertuis, D'Alembert and Lagrange at least \cite{Lanczos}, if not earlier.
In particular, a large number of works in Quantum Physics have used and continue to use as starting point, especially for calculational purposes, 
the stationary phase or saddle point approximation which rely on the vanishing variation under small perturbations of a judiciously chosen functional 
(the ``classical action" $\mathcal{S}$) \cite{Z-J}, an approach that can be traced back to an original idea of P.A.M. Dirac \cite{Dirac}. In this path-integral 
approach, as is very well-known, one starts with the path-integral/canonical partition function as the primary object encoding the statistically significant  properties 
of  the system
\begin{equation}   
        \mathcal{Z} \ = \ \int e^{-\mathcal{S}} \ [\mathcal{D}\phi ]
\end{equation}
where \ $\phi$ \ corresponds to the variables (``fields") in the action $\mathcal{S}$ to be integrated over and $[\mathcal{D} \phi ]$ represents an appropriate 
integration measure, which may not rigorously be known on whether it even exists, but whose ad hoc choice (usually being of Gaussian form) allows 
concrete calculations to be performed and eventually the results derived from it to be compared with experimental data. 
In the case of quantum gravity the most immediate choice for $\mathcal{S}$ is usually taken to 
be the Einstein-Hilbert or the Palatini actions, a Chern-Simons action, of their discretized counterparts etc, 
each of which may be augmented with boundary terms and/or topological terms etc.\\


\noindent{\small\bf Other entropies and robustness.} \ 
Before proceeding, we would like to have a short digression. During the last three decades, there have been several functionals that have been 
proposed purporting to capture the collective/thermodynamic  properties of systems of many degrees of freedom. One motivation for the formulation 
of such functionals, such as the ``Tsallis entropy" \cite{T-book}  or the ``$\kappa$-entropy" \cite{Kan} is to determine the thermodynamic properties 
of systems with long-range interactions. From the Newtonian viewpoint, gravity clearly falls in this category, as well as Maxwell's theory of electrodynamics etc. 
Assuming that such functionals may prove to be applicable  to a path-integral formulation of aspects of semi-classical or even quantum gravity, 
the arguments of the present work will still hold without any major modifications. 
The minor modification needed in case such functionals are pertinent, is to use in (70) another convex function instead of the exponential one, 
something akin to the aptly named ``$q$-exponential" \cite{T-book}, whose form will have to be determined. One could possibly use  the maximum entropy 
principle subject to appropriate constraints, for such a purpose. 
 A second minor, for our purposes, modification may be to substitute some other measure in the place of the often used  Gaussian measure in (70). 
Beyond these points, we expect the above analysis to still be valid. 
Another point that will most likely change is the rate of convergence to the limit of the saddle-point approximation, 
which is intimately related to the probability of dealing with space-time geometries that may be non-Euclidean. 
This in the spirit of theories having a statistical interpretation, where even when the ``classical" limit is known, it is the form of the 
``semi-classical" contributions/corrections that is used to distinguish between several competing models purporting to describe the same physical phenomenon. \\  


\noindent{\small\bf Generalized path integrals.} \ 
Going back to our argument,  in the spirit of the path-integral, an often discussed but still unsolved question is whether one should extend (70) by considering additional 
contributions by summing over  more ``primitive", than the metric, structures such over all topological, piecewise-linear, differentiable  etc. structures. 
Most of the treatments to quantum gravity that we are aware of, address the issue of a possible sum in the right-hand-side of (70) over all topologies. Then the modification of 
(70) states that the partition function of quantum gravity should be 
\begin{equation}
    \mathcal{Z}_{top} \ = \   \sum_{\mathrm{topologies}} \  \int e^{-\mathcal{S}} \ [\mathcal{D}\phi ]
 \end{equation}
Such possible summation over all topologies  has presented insurmountable 
difficulties, which can be credited in large part for the eventual demise of the dynamical triangulation approach to quantum gravity \cite{Ambj}, 
where one is faced with hard problems of G\"{o}delian type indecisive propositions.\\

On the other hand, (71) can be used as as starting point for a similar question, where the summation is not over all topologies of 4-dimensional manifolds, which are of 
course locally spaces endowed with a Euclidean metric. One could instead ask for summation over all metrics that can be placed on the underlying 
topological or uniform structure of the space whose classical limit will eventually be  a space-time. This however is would be a very broad set, 
hence a very difficult to analyze class of metrics. 
To be closer to something manageable, and also be in accordance with the equivalence principle demanding the local approximation of the 
underlying structure with linear spaces,  one may wish to consider locally, only  $p$-integrable Banach metrics, namely metrics/norms induced from $L^p(\mathbb{R}^n)$.
This effectively generalizes the underlying space-time structure from that of a Riemannian to something akin to a Finslerian space. 
Therefore one could write  a modified path-integral / canonical partition function, instead of (71), as 
\begin{equation}
     \mathcal{Z}_B \ = \   \sum_{p\geq 1} \ \int e^{-\mathcal{S}} \ [\mathcal{D}\phi ]
\end{equation}
with the summation being over all metrics of a space(-time) which are locally induced by the \ $L^p(\mathbb{R}^n)$. \ In conventional path-integral 
approaches to quantum gravity such a summation does not arise, since one has already determined to only use Euclidean metrics on the space-time underlying manifold.\\

 One could use the lack of renormalizability of the path-integral expressions like (70) around their saddle points, namely around  a fixed background metric,
 to argue against such an approach \cite{Velt}. After all, if (70) gives rise to non-renormalizable interactions around  Minkowski space, this should force us 
 to believe that it precludes (72) from  being more successful to that end.  Such a  criticism would be misguided though. Lack of renormalizability of (70) takes 
 into account only metrics of Riemannian (quadratic) form on the underlying space and the saddle point is calculated within the set of such metrics. What we 
 propose is to enlarge such a set to the more general locally $p$-integrable Banach metrics. Since our argument is quite empirical/``phenomenological" 
 rather than fundamental/dynamical, the issue of renormalizability of the underlying path-integral does not even enter our considerations. \\

Demanding a summation over such $p$-integrable metrics as in (72) may be reasonable, or not, but only after someone can properly write a finite classical action 
$\mathcal{A}$ for them. How exactly to do this is not clear to us at this stage. There are synthetic definitions of the Ricci and even the scalar curvature for 
metric measure spaces with very little regularity \cite{Vil}. One could use them and alongside a general minimal cost transportation to possibly argue 
in this direction. What is sorely lacking in such cases though is the formalism that could accommodate expressions for the non-gravitational fields capturing the essence
of the stress-energy tensor and recasting it in such a synthetic framework.   
Therefore a dynamical argument, which would be the most desirable,  in favor of Hilbert spaces and their induced Euclidean  metrics does not seem to be feasible, 
in any obvious way, at this stage.    \\


\noindent{\small\bf A ``kinematic" approach via smoothness.} \ 
Given  the above difficulties, we have to resort to ad hoc decisions in order to proceed. 
To the extent of our knowledge, there has never been  a variational principle of  ``maximum convexity" or a principle of ``minimal smoothness" that would single 
out Hilbert spaces among all Banach spaces. The path-integral / partition function approach to quantum Physics can be interpreted as suggesting that all 
allowed possibilities in quantum evolution should be considered in calculating quantities of interest, each possibility however being assigned with a different weight factor. 
Following this  viewpoint one can extend/stretch the domain of this interpretation to allow not only for a set of Riemannian metrics to contribute to the evolution of a 
gravitational system but also consider a broader class of possible metrics. To keep things close to the familiar territory of Riemannian/Lorentzian metrics we have used 
induced metrics on space-time locally induced by \ $L^p(\mathbb{R}^n)$, \ as was mentioned before. 
The familiar picture of space-time appears then as the classical limit of a theory of quantum gravity, and so are its associated properties like smoothness etc.\\
 
Given the assumed irregularity/granularity of space-time at a fundamental level
(expressed through spin networks in loop quantum gravity, partially ordered sets with discrete measures in the causal set approach, 
simplicial approximations and Regge calculus in causal dynamical triangulations etc) it may not be out of place to assume that nature 
chooses the least smooth class of metrics among  such induced metrics  from \  $L^p(\mathbb{R}^n), \ 1 \leq p < \infty$. \  
Therefore it is the Hilbert space metric/norm \ $L^2(\mathbb{R}^n)$ \ that provides the only  extremal, hence dominant, 
contribution in an ``extended" path-integral approach (72) which, in turn, induces its properties to the classical space-time limit of the quantum gravitational theory. 
In short, the induced metric of space-time inherits its Euclidean character from that of \ $L^2(\mathbb{R}^n)$ \ which dominates the 
path-integral (72), by being the least smooth.\\    
        
        
 \noindent{\small\bf Convexity and predictability.} \        
A somewhat complementary argument for Hilbert spaces and the induced Euclidean metric form  on space-time, can be made based on convexity and 
predictability.  As stated in the previous sections, the Hilbert spaces \  $\mathcal{H} = L^2 (\mathbb{R}^n)$ \ are the most convex among all \ $L^p (\mathbb{R}^n)$. \ 
Contrast the behavior of the norm/metric of \ $\mathcal{H}$ \ to those of the family of \ $L^p(\mathbb{R}^n)$ \ that are the least convex. 
 These are \ $L^1(\mathbb{R}^n)$ \ and  \ $L^\infty (\mathbb{R}^n)$ \ which are neither uniformly convex nor uniformly smooth, 
 nor are they reflexive, so they have been largely  ignored in most part of this work. 
 Nevertheless, use of these two spaces can help make this argument more transparent.\\
 
  Consider, for concreteness, the \ $L^1(\mathbb{R}^n)$ \  space, or to be more intuitive, the metric induced by the related $l^1$ norm on \ $\mathbb{R}^2$. \ 
 This metric, for \ $\vec{x} = (x_1, x_2), \ \vec{y} = (y_1, y_2)$ \  where the coordinates are considered with respect to a Cartesian system, is given by
\begin{equation}        
    d(\vec{x},\vec{y}) \ = \ |x_1-x_2| + |y_1-y_2|
\end{equation}
Then one can see that there is an infinity of geodesics connecting \  $\vec{x}$ \ and \ $\vec{y}$. \ This by itself is not a drawback: after all, the north and south poles of a 
sphere with the induced metric from the Euclidean space are also connected by infinite geodesics (the meridians). 
The definition of geodesics is not a problem either: in metric geometry \cite{BBI} 
they can be defined to be the isometric images of the unit interval. The problem exists because many of the geodesics between $\vec{x}$ 
and $\vec{y}$ in the (73) are branching: geodesics that have initially a common segment can separate after a while. If one assumes strong locality in a theory of gravity,
whose metric is even ``Euclidianized" (made positive-definite) after a Wick rotation, then this presents a problem with predictability. Since the theory 
does not possess any ``memory" in its formulation, how then one can make any prediction based on the behavior of geodesics which largely encode the 
underlying geometry, if the theory has branching geodesics? If, for instance, the action \ $\mathcal{S}$ \ or the resulting kinematic equations possessed some form 
``memory" as in the case of systems being modelled by fractional derivatives \cite{Laskin, Tarasov, Calgagni}, then the use of geometric structures with branching geodesics 
might not pose a serious problem to predictability. Knowing this, one may wish to stay as far away as possible from using metrics that may allow for the possibility of 
branching geodesics. It turns out that the Hilbert space \ $L^2(\mathbb{R}^n)$ \ is the furthest away from resembling \ $L^1(\mathbb{R}^n)$ \ which has branching geodesics, 
at least when one uses the modulus of convexity to quantify such a difference.\\

The counter-argument to the above is that \ $L^1(\mathbb{R}^n)$ \ has branching geodesics exactly because it is not uniformly convex. 
If someone chose any other \  $L^p(\mathbb{R}^n), \ p\neq 1$ \ then this problem would not exist. This is largely correct. 
However it assumes very much like many occasions in classical Physics that
some objects can be well-approximated by point particles, which in the absence of (non-gravitational) forces move along causal (in the Lorentzian signature framework)
geodesics. When the quantum nature of such an object comes into consideration though, even in a fixed, classical background space-time, this statement 
would not be accurate. The uncertaintly principle would prevent such point-like structures from existing; this introduces substantial  technical complications 
for any operator in the assumed Fock space of a quantum gravity theory following the canonical approach, 
such as loop gravity for instance: the operator has to be ``smeared", namely to 
act on test not at a particular point but on an appropriately chosen neighborhood of it before applying the canonical commutation relations \cite{Ashtekar}. \\

The result is that  a wave-function will sweep out a tube, for short times, rather than a line, in such a space-time as it evolves. 
For predictability purposes, since the Schr\"{o}dinger, the Klein-Gordon, the Dirac etc  equations involve usual derivatives, as 
opposed to being  integro-differential equations that may signify that 
memory effects are taken into account, it is quite important for such tubes not to have a branching property. 
Naively speaking and without getting into any details, we believe that this goal has the best chances of being realized, 
if the underlying space has a metric which is as far away from having branching geodesics as possible, 
which again brings us back to favoring the use  of a Hilbert space.\\    


\noindent{\small\bf The space-time metric from Hilbert spaces.} \ 
Going from \ $L^2(\mathbb{R}^n)$ \  to the metric of space-time itself is quite straightforward, in principle. 
Consider as the linear spaces of interest to be appropriate \ $A \subset \mathbb{R}^n$. \ 
For the case of point particles this will be the tangent to the particles's space-time evolution trajectory. 
We can then confine ourselves to the analysis of a  subspace of  \ $L^2(\mathbb{R}^n)$ \ which is comprised of the 
characteristic functions \  $\chi_A$ \ of such subspaces. Then the quadratic metric on \ $L^2(\mathbb{R}^n)$ \  gets 
induced on such $A$ which acquires itself a quadratic metric. Use the equivalence principle and ``patch together" such \  $A$ \  endowed with their 
Euclidean metrics  to form the space-time of interest. 
This is  a kinematic construction. The dynamics is provided by Einstein's equations, 
after a Wick rotation back to indefinite signature metrics. 
This transition between metrics of different signatures may involve several subtleties which may have to be 
addressed at that stage, but this is outside the scope of the present work.\\  
             

\section{Discussion and outlook.}

In this work we have presented a non-rigorous, conjectural argument that aims to explain why the metric of space-time, after being suitably Wick-rotated,  
obeys the Pythagorean theorem. Such a metric can be seen as descending from the  metrics of appropriate Banach spaces which provide 
a reasonable kinematical framework of a mesoscopic description / quasi-classical limit of models of quantum gravity. The advantage of this approach is also
its disadvantage: it is kinematical with ad hoc aspects. It does not attempt to delve into the actual realm of the approaches to quantum gravity, however it is motivated by and uses 
some of the common points of such approaches. We relied on aspects of General Relativity and the Einstein equations for some motivation, used the spirit of the 
variational principles and path-integrals to formulate our approach in the spirit of classical and quantum Physics and then for our proposal 
we  used standard results from functional analysis, convexity and the theory of normed spaces. \\ 

As in any,  partly ad hoc and conjectural,  work the virtues of the current work may be seen as out-weighting their shortcomings, or vice versa. 
The value of  work like the present, which may appear to purport to justify the ``obvious", may  lie in the ideas involved in it and in the methods 
used to reach the conclusions.  Much more importantly from a physical viewpoint, it may also point out to  reasons 
about the inapplicability of its conclusions should pertinent experiments refute our currently held ideas about space-time properties, 
or should one observe such exceptions at, or beyond,  the galaxy cluster or the Planck scales. \\  

We would like to add that there should be some skepticism regarding the role of the Wick rotation of the space-time signature to a positive-definite one. 
There is little doubt that  the Wick rotation has been very successful in obtaining results in perturbative quantum field theories, or more generally when employing 
saddle point approximations,  which may otherwise be ill-defined or  inaccessible in a covariant 
approach to such relativistic theories.  It is not clear, to us at least, to what extent employing such Euclidean (positive-definite) metrics  is equivalent to purely 
Lorentzian  arguments and results, in particular in regimes outside the possible domain of validity of such saddle-point approximations. 
It is well-known now that many fundamental results of Riemannian geometry (such as the Hopf-Rinow theorem etc) are either not valid in Lorentzian geometry
or may become valid when appropriate modifications are made.
Such discrepancies become even more pronounced when one considers the topological and causal formalism of space-times and other such statements of 
Lorentzian geometry which have no obvious analogue in the Riemannian case \cite{CGP, BEE, MingSan}.  
Since our results rely on positive-definite metrics of the underlying (linear) spaces, 
we can only be skeptical about their applicability in the physical, indefinite (Lorentzian) signature, case.\\

A technical point  that may be worth addressing before closing, is the fact that all the analysis in this work uses in a very essential way the particular moduli 
of convexity and smoothness whose definitions and properties have been stated above. Naturally, these are mathematical constructions and cannot conceivably 
be unique or automatically be considered  the ``most useful" among their peers.  And actually they are not. As mentioned in 
the opening paragraphs of this work, there are several, generally inequivalent, moduli encoding convexity and smoothness \cite{Fuster}. 
If history is any guide, new such moduli will keep being defined, their properties being examined and their values will be calculated in concrete cases in the future. 
We have chosen the above two moduli because they appear to us to be the simplest, 
the most intuitive and also the most developed. It is theoretically possible that other moduli may single out other spaces, as opposed to Hilbert ones, 
but we have not been able to find any such results in the existing literature nor can we see any overwhelming physical reason for their implementation.\\     

                                                       \vspace{1mm}


\noindent {\bf Acknowledgement:} This work was, largely, performed when the author was a faculty member at the Weill Cornell Medicine - Qatar. \\

                                                        \vspace{5mm}


\noindent{\bf Conflict of interest:} The author declares no conflict of interest.\\

                                                        \vspace{5mm} 
                                                        



\end{document}